\documentclass[lettersize,journal]{IEEEtran}

\usepackage{pifont}
\usepackage{calc}
\usepackage{comment}

\usepackage{psfig}
\usepackage{amsmath,amsfonts}
\usepackage{mdwtab}

\newcommand{\conjugn}{\mbox{$2^n \pm 1$}}
\newcommand{\fermn}{\mbox{$2^n+1$}}
\newcommand{\mersn}{\mbox{$2^n-1$}}

\usepackage{psfig}
\usepackage{comment}

\usepackage{textcomp}
\usepackage{dcolumn}
\usepackage{listings}
\usepackage{color}
\usepackage{xcolor}

\usepackage{pifont}
	\newcommand{\bi}{\begin{itemize}}
	\newcommand{\bii}{\begin{itemize}\item}
	
	\newcommand{\ei}{\end{itemize}}

\newtheorem{ttt}{Theorem}

\newtheorem{eee}{Example}

\newtheorem{ppp}{Property}

\newcommand{\mlabel}[1]{\mbox{\label{#1}}}
\newcommand{\mfig}[3]{%
\begin{figure}[!t]%
\centering 
\includegraphics[width=#2\columnwidth]{#1.eps}\caption{#3.}\mlabel{#1}%
\end{figure}}

\usepackage{color}

\usepackage{cite}
\usepackage[caption=false,font=normalsize,
   labelfont=sf,textfont=sf]{subfig}
\usepackage{textcomp}
\usepackage{stfloats}
\usepackage{url}
\usepackage{verbatim}
\usepackage{graphicx}
\usepackage{balance}

\begin{document}

\title{Efficient Implementations of Residue Generators %for Conjugate Moduli $2^n \pm 1$ 
Mod \fermn{} Providing Diminished-1 Representation}
\author{Stanis\l{}aw J. Piestrak \& Piotr~Patronik
\thanks{The authors are with 
Department of Computer Engineering, Faculty of Electronics (W‑4/K‑9), 
Wroc\l{}aw University of Science and Technology,
50-370 Wroc\l{}aw, Poland (email: stanislaw.piestrak@pwr.edu.pl; piotr.patronik@pwr.edu.pl).}
}

\maketitle

%Corresponding author
\markboth{%
%IEEE TRANSACTIONS ON COMPUTERS,
%Vol. xx, No. xx,  2025
}
{%PIESTRAK AND PATRONIK, EFFICIENT IMPLEMENTATIONS OF RESIDUE GENERATORS MOD \fermn
}

\begin{abstract}
The %conjugate moduli of the form \conjugn{} 
moduli of the form \fermn{} 
belong to a class of low-cost odd moduli,  
which have been frequently selected to form the basis of 
various residue number systems (RNS). 
%As for the modulus \fermn{}, 
The most efficient computations modulo (mod) \fermn{} are performed using
the so-called diminished-1 (D1) representation. 
Therefore, it is desirable that 
the input converter from the positional number system to RNS 
(composed of a set of residue generators) 
could generate the residues mod \fermn{} in D1 form. 
In this paper, we propose the basic architecture of the residue generator mod \fermn{} with D1 output. 
It is universal, 
because its initial part can be easily designed for an arbitrary $p \geq 4n$, 
whereas its final block---the 4-operand adder mod \fermn---preserves the same structure for any $p$. 
If a pair of conjugate moduli \conjugn{} belongs to the RNS moduli set, 
the latter architecture can be easily extended 
to build $p$-input bi-residue generators mod $2^n \pm 1$, 
which not only save hardware by sharing $p-4n$ full-adders, %capable of sharing hardware 
but also generate the residue mod \fermn{} directly in D1 form. %representation. 
%It is shown that all new designs enjoy the best performance 
%both in terms of hardware complexity and delay.  
\end{abstract}


%%%%%%%%%%%%%%%%%%%%%%%%%%%%%%%%%%%%%%%%%%%%%%%%%%%%%%%%%%%%%%%%%%
\section{Introduction}
%%%%%%%%%%%%%%%%%%%%%%%%%%%%%%%%%%%%%%%%%%%%%%%%%%%%%%%%%%%%%%%%%%
\IEEEPARstart{T}{he} residue arithmetic modulo (mod) \fermn{} has found numerous applications of which 
two distinct classes involve the non-positional residue number system (RNS) 
and the Fermat Number Transform (FNT) (where the Fermat number $F_n=2^{2^n} +1$). 
The RNS is defined by a set of pairwise prime natural numbers, called \emph{moduli}, 
whose product determines its \emph{dynamic range}. 
Its major attraction, which makes it competitive to the positional 2's complement system, 
is the possibility of particularly efficient 
(w.r.t. area, time, and power consumption) 
implementation of algorithms involving 
mostly sum of products operations. 
The numerous applications of RNS include implementations of the algorithms related to:
RSA public-key cryptosystem  \cite{Elango2023}, %\cite{Posch1992432a},\cite{Diel2023} 
%Montgomery multiplication \cite{Ahsan202272}, 
FIR filters  \cite{Piestrak2008127,Cardarilli2020186},  %,Balaji2024
%steganography \cite{Belhamra2020}, %\cite{Yao20181480 }
microprocessors \cite{Barraclough198924,Patronik20171031},
artificial intelligence \cite{Deng20241}, 
%neural networks \cite{Olsen20181}, 
%discrete wavelet transform (DWT) \cite{Alzaq20181}, %number-theoretic transforms
as well as many other %digital signal processing 
DSP 
applications \cite{Chang201526}. %\cite{Albicocco2014436}
%industrial applications (Position Estimators for Electrical Encoders) \cite{Cardarilli202398586} 
%motion estimation  \cite{Vayalil2019572} 
%
On the other hand, the FNT with diminished-1 representation was used 
to implement various DSP algorithms \cite{Daher2021103029}, %\cite{Agarwal197487,Agarwal1975550}
to accelerate integer convolutional neural networks \cite{Baozhou201947} 
and to reduce the computational complexity of the chromatic dispersion compensation 
in optical communication systems \cite{Xing20245190}.

In general, any set of pairwise prime moduli could be used to form an RNS. 
Nevertheless, some moduli like those of the type $2^k$ and 
the conjugate moduli of the form \conjugn{} 
have particularly hardware-efficient and fast implementations of the residue datapaths, 
and hence are called \emph{low-cost} moduli. 
%The conjugate moduli of the form \conjugn{} are low-cost odd moduli,  
%which have been frequently selected to form the basis of 
%various residue number systems (RNS). 
Because only one even modulus can be used, 
of particular interest are the remaining two classes, of which \mersn{} 
is unquestionably the best one. 
The next best class of odd moduli are those of the form \fermn{} for which, however, 
some problems have been identified and resolved as follows. 
The normal representation of residues mod \fermn{} 
requires $n+1$ bits to represent all valid values from $[0, 2^n]$,  
%More importantly, 
which means that only \fermn{} out of $2^{n+1}$ combinations are actually used. 
Amongst them, $(10 \ldots 0)$ is the 
only one out of $n+1$ combinations with the Most Significant Bit (MSB) set to 1. 
Leibowitz \cite{Leibowitz1976356} observed that 
%There has been a consensus that 
executing arithmetic operations on $(n+1)$-bit operands 
involves unnecessary hardware cost and delay. 
Assuming that any zero operand is recognized by a separate zero indication bit, 
the operations can be executed on $n$-bit operands, 
provided that each of them is decremented by 1. 
%whereas the operand equal to 0 is indicated by a separate bit. 
Such a notation was called the \emph{diminished-1 (D1) representation}. 
However, %the trade-off for using the D1 representation is the 
%potential cost of 
%to benefit from D1 representation,  that  
to make possible execution of cheaper D1 operations, 
a designer must ensure two following features. 

(i) The residue mod \fermn{} is provided to the datapath mod \fermn{} 
in D1 form. %, and %that 

(ii) Once the final result mod \fermn{} is available, 
either it is converted from the D1 form to the normal residue representation---% 
to make possible using generally available reverse converters, or 
it is applied directly to specially constructed reverse converter accepting the D1 from. 

%
%need for two conversions: 
%(i) from the normal representation --- to make possible execution of cheaper D1 operations, and 
%(ii) once the final result mod \fermn{} in D1 representation is available, 
%from the D1 representation to the normal representation --- 
%to make possible using generally available reverse converters. 
%Such converters have been proposed in the literature. 
%\mpp{where?}
%Obviously, the larger is the 
%%number of operations
%amount of computations 
%mod \fermn{} before executing the reverse translation, 
%the more advantageous is the D1 representation.

Because here we are particularly interested in the circuitry which involves the D1 representation, 
the following survey concentrates only on the contributions 
specifically taking into account the latter. 
%Some works presenting 
The most efficient implementations of the 
arithmetic circuits mod \fermn{} using  D1 representation are: 
adders \cite{Vergos20021389}, \cite{Vergos2010911},
multi-operand modulo adders (MOMAs) \cite{Vergos2010911}, 
multipliers \cite{Jaberipur201961}, and 
multiplier-accumulators (MACs---also called fused add-multiply units) 
\cite{Efstathiou2013410}, \cite{Tsoumanis20161}. 
%all for diminished-1 encoding 
The reverse converters for the special 3-moduli set %s like the 3-moduli set 
$\{ 2^n, 2^n-1, 2^n+1 \}$ and its extensions which include the fourth modulus $2^{2n}+1$, 
which were designed explicitly assuming that 
the datapath channels mod \fermn{} and $2^{2n}+1$ produce the D1 result, 
were proposed in \cite{Vassalos20131798}. 
An improved version of the reverse converter for the above 3-moduli set 
was proposed in \cite{Jaberipur201961} (Fig. 10). 

The input (forward) converter for any RNS-based processor essentially consists of 
a set of residue generators for all moduli defining an RNS. 
Here, we are interested in designing residue generators %which take into account two criteria: 
with two features: 
(1) they generate the D1 output for the \fermn{} modulus, and 
(2) they are amenable for hardware cost reduction by using shared logic 
with at least one residue generator for some other modulus.
Indeed, relatively little works can be found on this subject. 
The most obvious scheme of the residue generator mod \fermn{} which provides the output in D1 form 
consists of any normal residue generator mod \fermn{} 
(such as proposed in \cite{Piestrak199468})
followed by the converter into D1 form, 
e.g. in the form of the modified D1 adder connected to the CSA tree \cite{Vergos2010911} (Fig. 5b). 
The $3n$-bit input residue generator mod \fermn{} 
(intended only for the 3-moduli set $\{ 2^n, 2^n-1, 2^n+1 \}$) 
providing directly 
the D1 output was proposed in \cite{Jaberipur201961} (Figs. 4 and 8). 
%The $(3n+k)$-bit input residue generator mod \fermn{} ($0 \leq k \leq n$)
%(intended only for the 3-moduli set $\{ 2^{n+k}, 2^n-1, 2^n+1 \}$) 
%providing directly 
%the D1 output was proposed in \cite{Chaves2007472} (Fig. 6),
%
For some moduli sets, 
including those which contain the pair of the conjugate moduli like $2^n \pm 1$, 
%providing only weighted representation 
hardware savings can be obtained by using some shared circuitry to design input converters. 
The problem of hardware sharing between various residue generators 
for conjugate moduli $2^n \pm 1$
was considered in 
\cite{Pourbigharaz1994522,Piestrak20111435}.  %Skavantzos1999826,
%but only 
%In \cite{Pourbigharaz1994522}, it was considered in a limited context of designing 
%as a means to reduce the cost of binary-to-residue (input) converters 
%in some classes of RNS-based systems. 
However, 
these works deal only with the standard representation of residues mod \fermn. 
%Because, 
To our best knowledge, %as far as we know, 
no design methods of the multi-residue generators 
for conjugate moduli $2^n \pm 1$ using shared logic and providing 
diminished-1 representation have been reported yet. 
Obviously, the most evident (but not necessarily the most efficient) solution would be to use 
the bi-residue generator mod \conjugn{} from \cite{Piestrak20111435} 
%\cite{Pourbigharaz1994522,Skavantzos1999826,Piestrak20111435}
%a pair of separate residue generators mod \mersn{} and \fermn{} from \cite{Piestrak199468}, 
whose mod \fermn{} output feeds the normal-to-D1 converter, 
such as for example one from \cite{Vergos2010911}.

Therefore, 
the goal of this paper is to study the possibility 
of designing efficient residue generators mod \conjugn{}, 
which would produce directly the operand in D1 form for any number of input bits $p$ 
and would be easily amenable for hardware sharing with 
the mod \mersn{} residue generator.

%
%but only
%In \cite{Pourbigharaz1994522}, it was considered in a limited context of designing 
%as a means to reduce the cost of binary-to-residue (input) converters 
%in some classes of RNS-based systems. 

This paper is organized as follows.  
In Section \ref{sec-prelim}, some theoretical background on the D1 representation 
and the periodicity properties of the series of $|2^k|_A$ is summarized. 
In Section \ref{sec-gen-ferm-dim}, %{sec-bi-gen}, 
a new architecture of the universal residue generator mod \fermn{} 
providing the residue directly in D1 form, is detailed
along with
the possibility of its extension to the 
bi-residue generator mod \conjugn{}.
%is also presented. 
%In Section \ref{sec-eval}, 
%the gate level complexity evaluations of 
%new bi-residue generators and their existing counterparts is presented.
%Finally, some c
Conclusions are given in Section \ref{sec-concl}.

%\newpage
%%%%%%%%%%%%%%%%%%%%%%%%%%%%%%%%%%%%%%%%%%%%%%%%%%%%%%%%%%%%%%%%%%
\section{Preliminaries}\label{sec-prelim} %{Theoretical Background}
%%%%%%%%%%%%%%%%%%%%%%%%%%%%%%%%%%%%%%%%%%%%%%%%%%%%%%%%%%%%%%%%%%
%\vspace{-5mm}
%In this section, 
%we will present the basics on the diminished-1 (D1) representation 
%and the periodicity properties 
%of the series of $|2^k|_A$  %$2^k$ taken modulo any $A$ 
%and their applications in designing residue generators and MOMAs.
%
%as well as the details of the previous architectures of 2-residue generators 
%mod conjugate pairs \conjugn. 
%================================================================%
	\subsection{Modulo \fermn{} Diminished-1 (D1) Representation} %  Arithmetic
%================================================================%
The modulo \fermn{} diminished-1 (D1) representation, 
which was introduced in \cite{Leibowitz1976356},  
includes a zero indication bit. 
A number $X \in [ 0, \fermn )$ is represented as 
$X^* = (x_z\: x_{n-1} \ldots x_0)$, 
where $x_z$ is the zero indication bit and 
$X_{-1} = (x_{n-1} \ldots x_0)$ is the diminished-1 magnitude of $X$. 
Formally, the terms $x_z$ and $X_{-1}$ are defined as
\begin{equation}
x_z = \left\{ 
  \begin{array}{ c l }
0  & \quad \textrm{if }  X \neq 0 \\
1  & \quad \textrm{if }  X = 0 
\end{array}
\right.
\label{eqn-xz}\end{equation}
so that 
$X = \bar{x}_z + X_{-1}$.
%\begin{equation}
%X = \bar{x}_z + X_{-1}.
%\label{eqn-xdim1}\end{equation}

%================================================================%
	\subsection{Periodicity Properties of the Series of $|2^k|_{\conjugn}$}
%Numbers  
%$|2^k|_A$
%================================================================%
In the designs considered here, 
we will need the following notions, % and properties, 
which characterize the periodicity of the series of $\left| 2^k \right|_{\conjugn}$  
and are particularly useful to design efficient arithmetic circuits mod \conjugn{} 
\cite{Piestrak199468}. 
The practical importance of periodicity 
%stemming 
stems 
from the following equations, 
%Let $t$ denote a nonnegative integer. 
which hold for any nonnegative integer $j$: 
%was revealed in \cite{Piestrak1991100,Piestrak199468}: 
     \begin{equation} 
\left| 2^{jn+k} \right|_{\mersn} = \left| 2^k \right|_{\mersn} 
   \mlabel{period-mers}     \end{equation}
     \begin{equation} 
\left|2^{jn+k} \right|_{\fermn} = (-1)^j  \left|2^k \right|_{\fermn} .
   \mlabel{half-period-ferm}     \end{equation}
In particular, for $k=0$ the above equations take the form:
\begin{eqnarray}
\left| 2^{jn} \right|_{2^n-1} &&= 1
		\label{2jn-mod-mers}\\
\left| 2^{jn} \right|_{\fermn}&&= \left\{
\begin{array}{cl}
 1 & \ \mbox{if $j$ even}
\label{2jn-mod-ferm}\\
2^n = \left|-1\right|_{\fermn} & \ \mbox{if $j$ odd}
\end{array}
\right.
%\left| 2^{j\cdot 2n} \right|_{\fermn}&&= 1 \label{2j2n-mod-ferm}
\end{eqnarray}
In \cite{Piestrak199468}, 
it was shown how to exploit these properties to simplify 
designing %initially only %various circuits like 
residue generators and multi-operand modulo adders (MOMAs) 
by using carry-save adders (CSAs) with end-around carry (EAC). 
%The periodicity of the series of $\left| 2^i \right|_{A}$ and CSAs with EAC: 
%\cite{Piestrak199468}
In \cite{Piestrak199468}, 
it was shown that designing any arithmetic circuit taking advantage of 
Eqn \eqref{half-period-ferm} allows to 
invert all signals of weight $\left|2^{jn+k} \right|_{\fermn}$ for odd $j$ 
and handle them as signals of weight $\left|2^k \right|_{\fermn}$,  
provided that the correction constant equal to $|-2^k|_{\fermn}$ is added. 
To avoid unnecessary multiple additions of corrections, 
we can apply a simple general rule to calculate the cumulative correction value 
%$COR_{\fermn}$
for any arithmetic circuit taking advantage of Eqn \eqref{half-period-ferm}, 
%which 
%involves all inverted signals of weight $\left| 2^k \right|_{A}$, was 
given in \cite{Piestrak200231}: 
$COR_{\fermn}$ is obtained as the cumulative sum of all inverted signals of weight $\left| 2^k \right|_{\fermn}$ 
that appear in the circuit 
taken mod $\fermn$, 
which can be added at some stage of computation. 
%Obviously, 
In all circuits which will be considered here, 
the total correction $COR_{A}$ must be taken into account prior 
the final D1 representation is obtained.

%Eqns \eqref{mult-res-conjug-a} and \eqref{mult-res-conjug-b} 
Eqns \eqref{2jn-mod-mers} and \eqref{2jn-mod-ferm}
constitute the theoretical background 
for designing bi-residue generators for the conjugate moduli \conjugn.

%\newpage
%%%%%%%%%%%%%%%%%%%%%%%%%%%%%%%%%%%%%%%%%%%%%%%%%%%%%%%%%%%%%%%%%%
\section{Design of Residue Generators mod \fermn{} with D1 Output}\label{sec-gen-ferm-dim}
%%%%%%%%%%%%%%%%%%%%%%%%%%%%%%%%%%%%%%%%%%%%%%%%%%%%%%%%%%%%%%%%%%

%%%%%%%%%%%%%%%%%%%%%%%%%%%%%%%%%%%%%%%%%%%%%%%%%%%%%%%%%%%%%%%%%%
\subsection{Architecture Designed According to \cite{Piestrak199468}}
%%%%%%%%%%%%%%%%%%%%%%%%%%%%%%%%%%%%%%%%%%%%%%%%%%%%%%%%%%%%%%%%%%
Here, we will first present the design method of residue generators mod \fermn{} 
according to \cite{Piestrak199468} and then we will consider the possibilities to 
generate the output in D1 form. 

We assume that the input $p$-bit vector $X$ 
is sufficiently large, so that it can be partitioned into 
%     \begin{equation}
%r = \lceil p/n) \rceil > 2 
%\label{r-ferm}\end{equation}
%$n$-bit blocks. % $B_j$   
$r= \lceil p/n \rceil >2$ $n$-bit blocks $B_j$,  
beginning with the least significant bits (LSBs), 
i.e., $X = ( B_{r-1} \ldots B_1 B_0)$, where 
$B_0 = (x_{n-1} \ldots x_{1} x_{0})$, 
$B_1 = (x_{2n-1} \ldots x_{n+1} x_{n})$, 
etc. 
If $p$ does not divide $n$, 
the block $B_{r-1}$ containing %less than $n$ 
the most significant bits (MSBs) 
is padded with $r \cdot n - p$ leading 0s.

%Because $HP(\fermn)=n$ 
%for any $n \geq 2$ and  
Due to \eqref{half-period-ferm} we have 
%$|2^{j \cdot n}|_{\fermn} =(-1)^j$ for any $j\geq 0$, then \\[-3.5ex]
     \begin{subequations}\label{x-blocks-ferm} 
     \begin{align} %     \begin{multline}%     \begin{equation}  
\hspace*{-.9em}\left| X \right|_{\fermn} \!=\! 
\displaystyle\left| \displaystyle\sum_{j=0}^{r-1} 2^{j \cdot n} \!\cdot B_j\! \displaystyle\right|_{\fermn} 
\hspace*{-.5em}= 
\displaystyle\left| \displaystyle\sum_{j=0}^{r-1} (-1)^j B_j \displaystyle\right|_{\fermn}   
\label{x-blocks-ferm:a}\\
  = \displaystyle\left| \left( \sum_{j=0,\,j\,{\rm even}}^{r-1} B_j \right)
         \!-\! \left( \sum_{j=1,\,j\,{\rm odd}}^{r-1} B_j \right) \displaystyle\right|_{\fermn} .
     \label{x-blocks-ferm:b}   \end{align}%   \end{multline}%\end{equation}
\end{subequations}

%The decimal value mod \fermn{} of 
For any odd $j$, 
%$B_j = (b_{j,n-1} \ldots b_{j,1} b_{j,0})$ 
%is $\sum_{i=0}^{n-1} b_{j,i} 2^i$.  
%It can be easily shown that 
we can benefit from the following equality to replace the second part of Eqn \eqref{x-blocks-ferm:b} as follows
\begin{equation}
\Big| -B_j \Big|_{\fermn} = 
\Big| \bar{B_j} - \left( \sum_{i=0}^{n-1} b_{j,i} 2^i \right) \Big|_{\fermn} = 
\bar{B_j} +2 .
\label{eqn-notbj}\end{equation}
Eqn \eqref{eqn-notbj} indicates the correction constant equal to 2, 
which must be added mod \fermn{} to the final result. 

Basically, Eqns \eqref{x-blocks-ferm:b} and \eqref{eqn-notbj} 
can be used as a starting point to design the residue generator mod \fermn{} 
with D1 output. 
However, the following example reveals some problems involved with such a design approach. 
\begin{eee}
Consider the design of three residue generators mod 9 according to Eqn \eqref{x-blocks-ferm:b} for $p = \{ 16, 17 , 18 \}$. 
%We will show only 
Initially, 
the set of $p=18$ input bits is partitioned into $r= \lceil p/3 \rceil$ 3-bit blocks 
$B_0$, $B_1$, $B_2$, $B_3$, $B_4$, and $B_5$ (with padded two and one 0's, respectively for $p=16$ and $p=17$), 
in which all nonzero bits of the odd-numbered blocks $B_1$, $B_3$, and $B_5$ are complemented.  
To construct the CSA tree, 
the bits of the blocks $B_k$, $0 \leq k \leq 5$, are partitioned 
into $HP(9) = 3$ 
disjoint sets $G_k$, $0 \leq k \leq 2$, containing the bits of
the same weight $|2^k |_9$, i.e.:
\begin{eqnarray*}
G_0 &=& \{x_0, \bar{x}_3, x_6, \bar{x}_9, x_{12}, \bar{x}_{15} \}\\
G_1 &=& \{x_1, \bar{x}_4, x_7, \bar{x}_{10}, x_{13}, \bar{x}_{16} \}\\
G_2 &=& \{x_2, \bar{x}_5, x_8, \bar{x}_{11}, x_{14}, \bar{x}_{17} \} .
\end{eqnarray*}
(Obviously, for $p=16$ the bits $\bar{x}_{16}$ and $\bar{x}_{17}$ are omitted; 
similarly, for $p=17$ the bit $\bar{x}_{17}$ is omitted.) 
%We will show only 
The CSA parts of these residue generators can be  
described using the following shorthand notation 
introduced in \cite{Piestrak199468}. 
The contents of a column
$G_k$ alternately indicates either how many bits of residue weight
$|2^k|_9$ are present at a given stage of computation or specifies
the number of full-adders (FAs) and half-adders (HAs) that
operate on the bits from a given set $G_k$ (the current number
of such bits is provided by the entry in the same column in the
preceding row).

{\scriptsize%\small 
\vspace*{+2ex}
%\begin{center}
     (a) \begin{minipage}[t]{40mm}\begin{tabular}{|c|c|c|c} 
%\hline
\cline{1-3}
$G_2$ & $G_1$ & $G_0$ & \\
\cline{1-3}
\cline{1-3}
5 & 5 & 6 & \\
\cline{1-3}
FA HA & FA HA & 2 FAs & CSA Stage 1 \\
\cline{1-3}
4 & 4 & 4 & \\
\cline{1-3}
FA & FA & FA & CSA Stage 2 \\
\cline{1-3}
3 & 3 & 3 & \\
\cline{1-3}
FA & FA & FA & CSA Stage 3 \\
\cline{1-3}
2 & 2 & 2 & \\
\cline{1-3}
\end{tabular}\end{minipage}
%\end{center}
\vspace*{+2ex}

%\begin{center}
     (b) \begin{minipage}[t]{40mm}\begin{tabular}{|c|c|c|c} 
%\hline
\cline{1-3}
$G_2$ & $G_1$ & $G_0$ & \\
\cline{1-3}
\cline{1-3}
5 & 6 & 6 & \\
\cline{1-3}
FA HA & 2 FAs & 2 FAs & CSA Stage 1 \\
\cline{1-3}
4 & 4 & 4 & \\
\cline{1-3}
FA & FA & FA & CSA Stage 2 \\
\cline{1-3}
3 & 3 & 3 & \\
\cline{1-3}
FA & FA & FA & CSA Stage 3 \\
\cline{1-3}
2 & 2 & 2 & \\
\cline{1-3}
\end{tabular}\end{minipage}
%\end{center}
\vspace*{+2ex}

%\begin{center}
     (c) \begin{minipage}[t]{40mm}\begin{tabular}{|c|c|c|c} 
%\hline
\cline{1-3}
$G_2$ & $G_1$ & $G_0$ & \\
\cline{1-3}
\cline{1-3}
6 & 6 & 6 & \\
\cline{1-3}
2 FAs & 2 FAs & 2 FAs & CSA Stage 1 \\
\cline{1-3}
4 & 4 & 4 & \\
\cline{1-3}
FA & FA & FA & CSA Stage 2 \\
\cline{1-3}
3 & 3 & 3 & \\
\cline{1-3}
FA & FA & FA & CSA Stage 3 \\
\cline{1-3}
2 & 2 & 2 & \\
\cline{1-3}
\end{tabular}\end{minipage}
}
%\end{center}
\vspace*{-2ex}
\begin{figure}[htb]%
\caption{Shorthand notation of the CSA tree for the residue generator mod 9 with: 
(a) $p=16$; (b) $p=17$; and (c) $p=18$ inputs.}
\label{fig-csa-16-17-18-mod9}\end{figure}

%Because the blocks $B_1$, $B_3$, and $B_3$

\begin{table}[!t]\begin{center}\caption{The corrections required due to complemented signals.}\label{tab-cor16-17-18}

\begin{tabular}{|c||c|c|c|c||c|}
\hline
$p$ & $B_1$ & $B_3$ & $B_5$ & CSA Stages 1--3 &  $COR(p, 9)$\\
\hline\hline
16 & $-7$ & $-7$ & $-1$ & $-2-1-1$ & $|-19|_9 = 8$ \\
\hline
17 & $-7$ & $-7$ & $-3$ & $-2-1-1$ & $|-21|_9 = 6$ \\
\hline
18 & $-7$ & $-7$ & $-7$ & $-2-1-1$ & $|-25|_9 = 2$ \\
\hline
\end{tabular}\end{center}\end{table}

In Fig. \ref{fig-csa-16-17-18-mod9}, 
it is seen that the CSA trees are virtually identical for the three values of $p$ 
with four identical inverted EACs 
(two for Stage 2 and one for Stage 2 and 3); the only difference results from  
one or two HAs replacing FAs for $p=17$ and $p=16$, respectively. 
% one for $p=16$ tit suffices to replace with two FAs from CSA Stage 1 minor differences
%The general rule to calculate $COR(p, \fermn)$
%
The final column of Table~\ref{tab-cor16-17-18} shows that, 
despite that the CSA tree reduces the input bits to the same set of six equally distributed bits, 
in each case the nonzero correction that must be added by the final adder mod \fermn{} differs. 
%As a result, 
Consequently, 
the final adder which generates the residue mod \fermn{} in D1 form must be adapted 
to include the correction depending on the number of inputs $p$. 
\IEEEQED
%(Nevertheless, for instance, it can be easily shown that e.g. for $18 \leq p \leq 21$ the correction is the same, NOOOO!
%because the block $B_6$ contains no complemented bits; BUT there are extra complemented EACs.)
\label{ex-16-17-18mod9}\end{eee} 
%\newpage

%%%%%%%%%%%%%%%%%%%%%%%%%%%%%%%%%%%%%%%%%%%%%%%%%%%%%%%%%%%%%%%%%%
\subsection{New Universal Architecture}
%%%%%%%%%%%%%%%%%%%%%%%%%%%%%%%%%%%%%%%%%%%%%%%%%%%%%%%%%%%%%%%%%%
The alternative new architecture of the residue generators mod \fermn{} proposed here 
will not have the previously indicated drawback. 
%We will propose the new design method of the residue generators mod \fermn{}
Moreover, besides generating the D1 output for any arbitrary $p$ 
without the need to add any correction due to complemented signals, 
%which, in advance, takes 
it will also take into account the possibility of hardware sharing 
with the residue generator mod \mersn. 
The latter relies on using two following equations:
     \begin{eqnarray}
\Big| \left| X \right|_{2^{2n}-1} \Big|_{\mersn}  & = & \left| X \right|_{\mersn}
\label{mult-res-conjug-a}\\
\Big| \left| X \right|_{2^{2n}-1} \Big|_{\fermn} & = & \left| X \right|_{\fermn},
 \label{mult-res-conjug-b}
\end{eqnarray}
which are the special cases of the well-known identity 
\begin{eqnarray}
\left| a \right|_{b} & = & \left| \left| a \right|_{bc} \right|_{b} .
		\label{eq-modbc-modb}
\end{eqnarray}

We assume that the input $p$-bit vector $X$ is partitioned into 
$q= \lceil p/(2n) \rceil$ $2n$-bit blocks $D_j$,  
beginning with the LSBs, %least significant bits (LSBs), 
i.e., $X = ( D_{q-1} \ldots D_1 D_0)$, where 
$D_0 = (x_{2n-1} \ldots x_{1} x_{0})$, 
$D_1 = (x_{4n-1} \ldots x_{2n+1} x_{2n})$, 
etc. 
If $p$ does not divide $n$, 
the block $D_{q-1}$ containing %less than $n$ 
the MSBs %most significant bits (MSBs) 
is padded with $q \cdot n - p$ leading 0s. 
%Initially, 
We assume that $p$ is sufficiently large, so that $X$ can be partitioned into 
     \begin{equation}
q = \lceil p/2n) \rceil \geq 4 
\label{q-2m}\end{equation}
$2n$-bit blocks. % $D_j$:   
%%%%%%%%%%%%%%%%%%%%%%%%%%%%%%%%%%%%%%%%%%%%%%%%%%%%%%%%%%%%%%%%%%
%The special cases of $q < 4$ will be considered later separately at the end of this section.
%\mpp{DO IT}
%%%%%%%%%%%%%%%%%%%%%%%%%%%%%%%%%%%%%%%%%%%%%%%%%%%%%%%%%%%%%%%%%%

%Initially, 
First, 
the $q$-operand $2n$-bit CSA with EAC  
reduces $p$ input bits to a pair of $2n$-bit vectors $D_C$ and $D_S$ 
by realizing the equation  
\begin{equation}
\left|	\sum_{j=0}^{s-1} D_j \right|_{2^{2n}-1} = \left| D_C + D_S \right|_{2^{2n}-1} .
\label{sum-d-2n}\end{equation}
Now each of thus obtained $2n$-bit vectors $D_C$ and $D_S$ 
can be split into a pair of $n$-bit groups 
containing $n$ MSBs and LSBs denoted, respectively, by the indexes $H$ and $L$: 
$D_C = (D_{C,H} \| D_{C,L} )$ and $D_S = (D_{S,H} \| D_{S,L} )$.  
By taking into account that  
%\begin{equation}
$D_C  = 2^nD_{C,H} + D_{C,L}$,  
%\label{chcl}\end{equation}
$D_S  = 2^nD_{S,H} + D_{S,L}$,  
and 
$\left| 2^n \right|_{\fermn} = \left| -1 \right|_{\fermn}$, 
we obtain the identity
\begin{eqnarray}
\left| X \right|_{\fermn} 
	& = & \Big|\left| D_C + D_S \right|_{2^{2n}-1}\Big|_{\fermn} 
		\nonumber %\label{xcs}
	\\
	& = & \Big| D_C + D_S \Big|_{\fermn} \nonumber
	\\
	& = &  \Big| (D_{C,H} \| D_{C,L}) + (D_{S,H} \| D_{S,L}) \Big|_{\fermn} \nonumber
	\\
	& = &
	\Big| 2^nD_{C,H} + D_{C,L} + 2^nD_{S,H}  + D_{S,L}  \Big|_{\fermn} \nonumber %\label{2nch-cl-2nsh-sl}
	\\
	& = &
	\Big| -D_{C,H} + D_{C,L} - D_{S,H}  + D_{S,L} \Big|_{\fermn} \label{ch-cl-sh-sl} .
		\end{eqnarray}
Eqn \eqref{ch-cl-sh-sl} can be realized using the 4-operand CSA mod \fermn{} 
followed by the special adder mod \fermn{} 
%to yield $|X|_{\fermn}$ in D1 form. 
to obtain the $(n+1)$-bit vector $X^*$, 
which appears directly in D1 form 
(see the proof of Eqn \eqref{xdim-fin} given below). 
(As the final adder mod \fermn{} with D1 output 
can be used one described in \cite{Zimmermann1999158}, 
whose detailed structure can be found on Fig. 7 in \cite{Vergos20021389}.) 
Figure \ref{new-genmod-dim1} shows 
the internal structure of the new residue generator mod \fermn{} with D1 output, 
designed according to the above procedure. 
Obviously, for $p=4n$, 
the whole circuits reduces to %it is nothing else but 
the final 4-operand adder mod \fermn.

\mfig{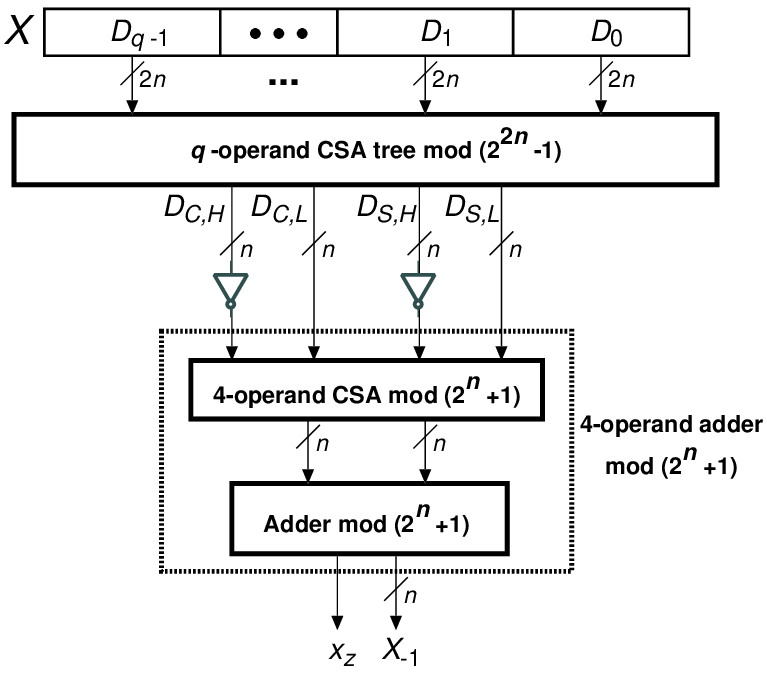}{0.7}{%Internal structure of the 
New residue generator mod \fermn{} with D1 output}

%%%%%%%%%%%%%%%%%%%%%%%%%%%%

%%%%%%%%%%%%%%%%%%%%%%%%%%%%

The theorem given below will prove that the output produced by the circuit of Fig. \ref{new-genmod-dim1} 
is in the D1 form indeed. 
However, besides the following identity ($a\in\{0, 1\}$)
\begin{equation}
- a = \bar a-1, 
\label{nota}\end{equation} 
we will need the following properties. % provided here with their proofs.

\begin{ppp}
For three natural numbers $0 \leq x, y, z <2^n$, the following equation holds:
\begin{eqnarray}
	&&
		\hspace{-19mm}
	\left| x+y+z \right|_{\fermn}
		\nonumber
	\\
	& \overset{\text{\scriptsize CSA}}{=} &
\left| 2c+s \right|_{\fermn} \nonumber\\
	&=&
\left| 2\cdot2^{n-1}c_{n-1} + 2(c_{n-2}\ldots c_0)  +s \right|_{\fermn} \nonumber \\
	&=&
\left| 2^nc_{n-1} + 2(c_{n-2}\ldots c_0)  +s \right|_{\fermn} \nonumber \\
	&=&
\left| 2(c_{n-2}\ldots c_0) -c_{n-1}  +s \right|_{\fermn} \nonumber \\
	&=&
\left| 2(c_{n-2}\ldots c_0) +\bar c_{n-1}-1  +s \right|_{\fermn} \nonumber	\\
	&=&
\left| (c_{n-2}\ldots c_0 \|c_{n-1})-1  +s \right|_{\fermn}
		\label{csa-ferm}
\end{eqnarray}
\label{prop-csa-ferm}\end{ppp} 
\vspace*{-3ex}

\begin{ppp}
For three natural numbers $ 0 \leq x, y <2^n$ and $0 \leq t \leq 2^n$, the following equation holds: 
\begin{eqnarray}
\left| x+y \right|_{\fermn}
	& = & \left| 2^nc_{n-1} + s \right|_{\fermn} \nonumber \\ 
	& = & \left| s-c_{n-1} \right|_{\fermn} \nonumber \\
	& = & \left| s + \bar c_{n-1} -1 \right|_{\fermn}	\nonumber \\
	& = & \left| t -1 \right|_{\fermn} .
		\label{cpa-dim1}\end{eqnarray}
\label{prop-cpa-dim1}\end{ppp} 
\vspace*{-3ex}

Property \ref{prop-csa-ferm} will be used twice to justify the computations 
executed by two subsequent CSA stages, 
whereas Property \ref{prop-cpa-dim1} will be used to evaluate 
the final result provided by the special version of the 
final adder mod \fermn.

\begin{ttt}
For the circuit of Fig. \ref{new-genmod-dim1} the following two equations hold:
\begin{eqnarray}
\Big| X \Big|_{\fermn} & = & \Big| D_{C,L} - D_{C,H} + D_{S,L} - D_{S,H} \Big|_{\fermn} \nonumber \\
& = & \Big|X^* +1\Big|_{\fermn} 
	\label{xdim-fin}\\
 & \mbox{and} & \nonumber \\
\left| X -1 \right|_{\fermn} & = & X^* .\label{xstar}
\end{eqnarray}
\label{theorem-dim1}\end{ttt}
%\begin{proof}
\vspace*{-4ex}
\noindent\emph{Proof.}
%First, notice that from Eqn \eqref{eqn-notbj} we have
Here, we will use the following identities:
$|- D_{C,H}|_{\fermn} = |\bar D_{C,H} +2|_{\fermn}$ and $|- D_{S,H}|_{\fermn} = |\bar D_{S,H} +2|_{\fermn}$.
Then
\begin{eqnarray}
	&& \hspace{-15mm} \Big| X \Big|_{\fermn} 
	\nonumber
\\
& \overset{\text{\scriptsize \eqref{ch-cl-sh-sl}}}{=} &
%& \overset{\text{\scriptsize \eqref{sum-d-2n}}}{=} &
\Big|	D_{C,L} - D_{C,H} + D_{S,L} - D_{S,H} \Big|_{\fermn} 
\nonumber
\\
%& =& \Big|D_{C,L} + \bar D_{C,H} -(2^n-1) + D_{S,L} + \bar D_{S,H} -(2^n-1)\Big|_{\fermn} 
%\nonumber \\
& =&
\Big|D_{C,L} + \bar D_{C,H} +2 + D_{S,L} + \bar D_{S,H} +2	\Big|_{\fermn} 
%	\label{cpa-dim1a}
\nonumber
\\
& \overset{\text{\scriptsize \eqref{csa-ferm}}}{=} &
\Big|D_{C,1} + D_{C,2} -1 +2 + \bar D_{S,H} +2\Big|_{\fermn} \nonumber
\\
& = &
\Big|D_{C,1} + D_{C,2} + \bar D_{S,H} +3\Big|_{\fermn} \nonumber
\\
& \overset{\text{\scriptsize \eqref{csa-ferm}}}{=} &
\Big|D_{C,3} + D_{C,4} -1 +3\Big|_{\fermn} \nonumber
\\
& = &
\Big|D_{C,3} + D_{C,4} +2\Big|_{\fermn} 
	\nonumber %\label{cpa-dim1b}
\\
& \overset{\text{\scriptsize \eqref{cpa-dim1}}}{=} &
\Big|X^* -1 +2	\Big|_{\fermn} \nonumber
\\
& = &
\Big|X^* +1\Big|_{\fermn} 
\hspace*{-3em}\label{xdim-proof}
\end{eqnarray}
By subtracting 1 from both sides of Eqn \eqref{xdim-proof}, we obtain 
\begin{eqnarray}
%\Big| X \Big|_{\fermn} 
%	& \overset{\text{\scriptsize \eqref{xdim-fin}}}{=} &
%	\Big|X^* +1\Big|_{\fermn} \nonumber	\\
\Big| X -1 \Big|_{\fermn} 
%	& \overset{\text{\scriptsize \eqref{xdim-fin}}}{=} &
	& = &
	\Big|	X^* +1-1 \Big|_{\fermn} \nonumber
	\\
	& = &
	\Big|X^* \Big|_{\fermn}  = X^* . \nonumber~~~\IEEEQED %\label{xstar}
\end{eqnarray}
%the latter leads to the final conclusion
%\begin{equation}
%	X^* \overset{\text{\scriptsize \eqref{xstar}}}{=} \Big| X -1 \Big|_{\fermn} .
%\label{dim1-final}\end{equation}
%
%\end{proof}
%\hfill$\square$

We have shown that for any $p$, $COR(\fermn, p)=0$, i.e., 
no correction needs to be added to obtain the D1 output. 
This is because the initial $s$-operand CSA tree mod $(2^{2n} -1)$ 
contains no inverted signals, 
whereas the rest of the circuit remains identical for any $p$.

%%%%%%%%%%%%%%%%%%%%%%%%%%%%%%%%%%%%%%%%%%%%%%%%%%%%%%%%%%%%%%%%%%
%\subsection{The Case of $p \leq 4a$}
%%%%%%%%%%%%%%%%%%%%%%%%%%%%%%%%%%%%%%%%%%%%%%%%%%%%%%%%%%%%%%%%%%
%
%For $p=4a$, it suffices to ignore the input $q$-operand CSA tree mod $2^{2n}-1$, 
%and the final result still appears in the D1 form.  

%%%%%%%%%%%%%%%%%%%%%%%%%%%%%%%%%%%%%%%%%%%%%%%%%%%%%%%%%%%%%%%%%%
\subsection{Design of Bi-residue Generators mod \conjugn}
%%%%%%%%%%%%%%%%%%%%%%%%%%%%%%%%%%%%%%%%%%%%%%%%%%%%%%%%%%%%%%%%%%

The architecture of the bi-residue generator mod \conjugn{} 
results from the following straightforward application of Eqns 
\eqref{mult-res-conjug-a} and \eqref{sum-d-2n} 
\begin{eqnarray}
\left| X \right|_{\mersn} 
	& = & \Big|\left| D_C + D_S \right|_{2^{2n}-1}\Big|_{\mersn} 
		\nonumber %\label{xcs}
	\\
	& = & \Big| D_C + D_S \Big|_{\mersn} \nonumber
	\\
	& = &  \Big| (D_{C,H} \| D_{C,L}) + (D_{S,H} \| D_{S,L}) \Big|_{\mersn} \nonumber
	\\
	& = &
	\Big| 2^nD_{C,H} + D_{C,L} + 2^nD_{S,H}  + D_{S,L}  \Big|_{\mersn} \nonumber %\label{2nch-cl-2nsh-sl}
	\\
	& = &
	\Big| D_{C,H} + D_{C,L} + D_{S,H}  + D_{S,L} \Big|_{\mersn} \label{ch-pcl-psh-psl} .
		\end{eqnarray}
Obviously, the $q$-operand $2n$-bit CSA with EAC,   
which reduces $p$ input bits to a pair of $2n$-bit vectors $D_C$ and $D_S$ 
according to Eqn \eqref{sum-d-2n} can be shared. 
The detailed internal structure of thus obtained %\fermn{} part of the sample 
bi-residue generator mod \conjugn{} 
%Fig. \ref{bi-residue-architectures}(c) 
is shown in Fig. \ref{bi-residue-conjug-dim1}.
%In the general case, 
%The upper part of the circuit of Fig. \ref{bi-residue-conjug-dim1} 
Its upper part allows to save $p-4n$ full-adders (FAs). 

\mfig{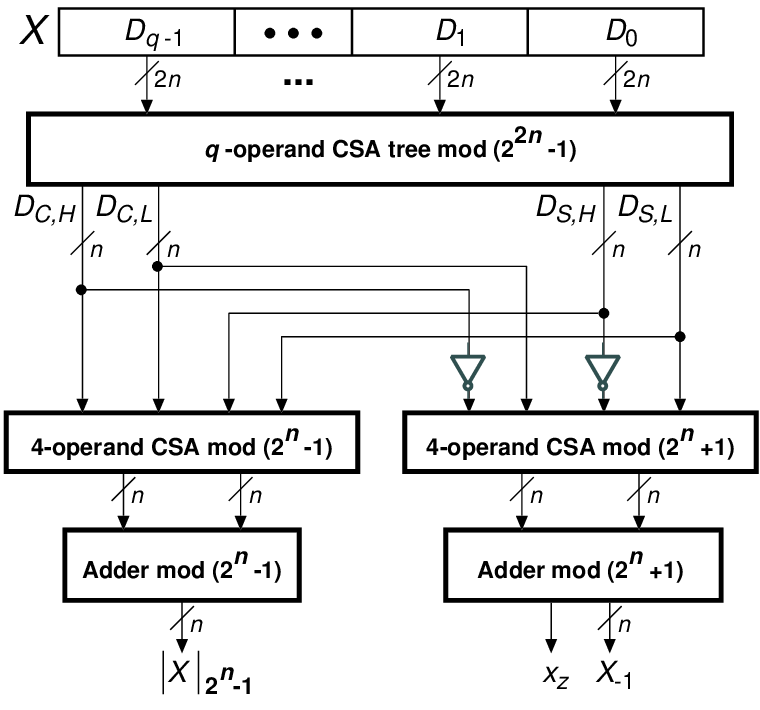}{0.7}{%Internal structure of the 
New bi-residue generator mod \conjugn{} with D1 output}

%The differences between various architectures of the input converters 
%handling the pair of conjugate moduli \conjugn{} 
%can be easily seen by inspection of Fig. \ref{bi-residue-architectures}. 
%
%\mfig{bi-residue-architectures}{0.4}{Three basic architectures of the input converters handling the pair of conjugate moduli \conjugn{}: (a); (b); and (c)}
%
%(a) A pair of independent residue generators mod \mersn{} and \fermn{}, 
%%with the converter to D1, added to the latter; 
%where the latter is followed by the special converter to D1, 
%designed e.g. according to %\cite{Zimmerman1999158} or 
%\cite{Vergos20021389}. 
%
%(b) A pair of independent residue generators mod \mersn{} and \fermn{}, 
%where the converter to D1 is integrated in the latter.
%% contains an embedded. 
%
%(c) The bi-residue generator mod \conjugn{} with shared $p$-bit CSA mod $2^{2n}-1$ proposed here, 
%where the final adder mod \fermn{} by construction generates the D1 output. 
%%converter to D1 is integrated in the final part of the residue generator. 

%%%%%%%%%%%%%%%%%%%%%%%%%%%%%%%%%%%%%%%%%%%%%%%%%%%%%%%%%%%%%%%%%%
\section{Conclusions}\mlabel{sec-concl}
%%%%%%%%%%%%%%%%%%%%%%%%%%%%%%%%%%%%%%%%%%%%%%%%%%%%%%%%%%%%%%%%%%
The diminished-1 (D1) encoding has been known for several years as the efficient approach 
which could improve performance of residue arithmetic circuitry modulo \fermn{} in 
arithmetic units using RNS. 
In this paper, 
we have proposed the new architecture of 
the $p$-input residue generator mod \fermn{} with D1 output. 
It can be useful to build an input converter for any RNS moduli set containing one or more moduli of the form \fermn. 
The circuit is universal, 
because its initial part can be easily designed for an arbitrary $p \geq 4n$, 
whereas its final block---the 4-operand adder mod \fermn---preserves the same structure for any $p$. 
The latter feature was shown essential for the possible easy extension  
to build $p$-input bi-residue generators mod $2^n \pm 1$ with shared logic, 
which allows to save $p-4n$ full-adders.  
The latter design can be useful for any 
%set of RNS moduli containing at least one modulus of the type \fermn{}  
%and for the latter for any 
set of RNS moduli containing a pair of conjugate moduli $2^n \pm 1$.
%
%The input converter for a pair of conjugate moduli \conjugn{}, 
%which uses shared logic and generates the residue 
%mod \fermn{} using the diminished-1 encoding, 
%was presented for the first time. 
%
%In summary, 
As far as we know, 
to date no general design methods of residue generators mod \fermn{} using shared logic 
with an arbitrary number of inputs $p$ 
and providing the input in D1 form have been presented yet. 
%The parameter estimation suggests that the proposed circuitry is not only 
%the least complex but also the fastest currently known.

%\newpage
\bibliographystyle{IEEEtran.bst}
\bibliography{rns,other} 
%\bibliography{rns,other,dim1} 

\end{document}